\documentclass[runningheads]{llncs}
\usepackage[T1]{fontenc}
\usepackage{graphicx}
\begin{document}
\title{To Patch or Not to Patch: Motivations, Challenges, and Implications for Cybersecurity}

\titlerunning{Incentives and Disincentives to Patching}

%
\author{Jason R.C. Nurse \orcidID{0000-0003-4118-1680}}
\authorrunning{J.R.C. Nurse}
%
\institute{Institute of Cyber Security for Society and School of Computing, \\University of Kent, Kent, UK \\
\email{j.r.c.nurse@kent.ac.uk}}
\maketitle              
\begin{abstract}

As technology has become more embedded into our society, the security of modern-day systems is paramount. One topic which is constantly  under discussion is that of patching, or more specifically, the installation of updates that remediate security vulnerabilities in software or hardware systems. This continued deliberation is motivated by complexities involved with patching; in particular, the various incentives and disincentives for organizations and their cybersecurity teams when deciding whether to patch. In this paper, we take a fresh look at the question of patching and critically explore why organizations and IT/security teams choose to patch or decide against it (either explicitly or due to inaction). We tackle this question by aggregating and synthesizing prominent research and industry literature on the incentives and disincentives for patching, specifically considering the human aspects in the context of these motives. Through this research, this study identifies key motivators such as organizational needs, the IT/security team's relationship with vendors, and legal and regulatory requirements placed on the business and its staff. There are also numerous significant reasons discovered for why the decision is taken not to patch, including limited resources (e.g., person-power), challenges with manual patch management tasks, human error, bad patches, unreliable patch management tools, and the perception that related vulnerabilities would not be exploited. These disincentives, in combination with the motivators above, highlight the difficult balance that organizations and their security teams need to maintain on a daily basis. Finally, we conclude by discussing implications of these findings and important future considerations.

\end{abstract}

\section{Introduction}
Cyber-attacks can originate from any location across the world and impact businesses, organizations, governments, and individuals. Ransomware, for instance, is currently one of the most significant threats and a key reason for its prominence is the challenge of identifying, capturing, and prosecuting geographically dispersed attackers~\cite{maccoll2024ransomware,mott2023between}. Another reason why threats such as ransomware, remote code execution and attacks on cloud environments are increasing in prominence is the fact that a large number of technology systems possess vulnerabilities~\cite{cisa2025cat,sowka2025towards}. Disclosed vulnerabilities have increased significantly over time with several reasons cited, including: strong competition by technology vendors leading to shorter product development and testing lifecycles; the large amount of outdated and legacy systems still continuously in use; a lack of regulation and legislation pertaining to vulnerabilities and software defects; more complex systems (e.g., the Internet of Things and cyber-physical system (CPS) advances~\cite{perez2024patchy}); and malevolent actors investing more resources into vulnerability research~\cite{ENISA2018sw}.

While existing but yet to be discovered vulnerabilities (zero-days) clearly pose a challenge, known (and disclosed) vulnerabilities also raise several issues. In an ideal world, as vulnerabilities are disclosed, product vendors and software developers would notify those who possess the systems and provide timely patches. Patches are defined as updates that act to remediate security vulnerabilities or fix bugs or other errors in a system. The patching of systems has become such a significant event and undertaking that in order to allow organizations across the world to prepare for updates, Microsoft (and others) instituted ‘Patch Tuesday’ and it has been one of the most anticipated days for security and IT professionals each month. 

The challenge, however, is that we do not live in an ideal world and often patches issued to system and product owners (be they in business, government, or are members of the public) are not implemented. To take the example of the WannaCry ransomware attack of 2017; this attack resulted in reports of at least 45 UK NHS hospital groups unable to use their IT systems and having to cancel appointments, surgeries, and divert patients~\cite{NAO2017hos}. A subsequent National Audit Office investigation into the attack found that it could have been prevented – or at least its impact minimised – had organizations implemented “basic IT security best practice”, including activities such as patching IT systems as directed by NHS Digital~\cite{NAO2017hos}. This failure to act on security patches is not isolated and can be seen in various other incidents including the substantial data breach at Equifax in 2017~\cite{USS2017eq}, and the concerns regarding the Log4j vulnerability in 2021 and MOVEit vulnerability in 2023~\cite{NCSC2024vul}. In fact, new research has found that 25\% of high-risk Common Vulnerabilities and Exposures (CVEs) are exploited on the day of their publication~\cite{Abbasi2023vul}.

A pertinent question therefore is, why do organizations and cybersecurity teams fail to patch when notified given it is often in their interest to protect themselves as best as possible. One reason frequently quoted is that of the potential adverse effects, a particularly salient point after the recent CrowdStrike incident where a defective patch resulted in widespread system outages~\cite{bbc2024crowd}. In research and practice, there have been several articles exploring the incentives for patching as well as the disincentives. The difficulty nevertheless is that this work scattered across academic studies, industry reports, and government documents, and it rarely considers the human aspects perspective. In this paper, we seek to tackle this challenge by aggregating and synthesizing prominent research in the field to provide a topical review on the incentives and disincentives for patching. This critical review pays special attention to the human aspects perspective and how it impacts these motives. The remainder of this paper is structured as follows: Section~\ref{sec:background} provides a brief overview of pertinent background on patching and its importance. Section~\ref{sec:incentives} explores the incentives for patching, including practices and motives which cause organizations to patch their infrastructure. In Section~\ref{sec:disincentives}, we concentrate on the disincentives for patching and identify primary factors which lead to organizations not updating their systems or failing to do so in a timely fashion. Finally, Section~\ref{sec:reflection} reflects on the previous sections, their implications and important future considerations, while Section~\ref{sec:conclusion} concludes this article.

\section{Patching: Nature and Context}
\label{sec:background}

Various entities (or groups of individuals) are involved in the identification, disclosure, and remediation of vulnerabilities. First, there is the product vendor or developer; this group is responsible for creating the product and issuing updates as vulnerabilities are identified. Next is the group involved in identifying vulnerabilities, for instance, these may be sanctioned security researchers (e.g., internal to vendors, or those emerging through bug bounty or vulnerability disclosure programs) or malicious actors (e.g., nation states or cybercriminal gangs). Product owners, are the final primary group and are those who use a vendor's products and are responsible for implementing any patches issued. In an organization, this last group may be deconstructed even further to consider product owners and the employees who actually use the product; for example, an employee who uses cybersecurity software (e.g., a VPN) purchased by their organization (i.e., the product owner). Frei et al. provide a succinct overview of the interaction between these groups including the various possible pathways from vulnerability discovery to disclosure, patching and exploitation~\cite{frei2010modeling}. 

A topic central to understanding why organizations -- and particularly, their IT/security staff -- apply or fail to apply patches is the patch management lifecycle itself. This involves identifying vulnerable technologies within the organization, deploying and installing the patches, and ongoing monitoring for patch compliance. While there are several ways to present this process, Dissanayake et al.~\cite{dissanayake2022software} suitably contextualize it within the overall vulnerability and patch timeline. They define essential tasks for security and IT teams as patch information retrieval, vulnerability scanning, assessment and prioritization, patch testing, patch deployment, and post-deployment patch verification. Each of these activities is critical for organizations to perform and require IT and security staff to have a comprehensive inventory of their systems, tools to scan for the identified vulnerabilities, environments to test patches prior to deployment, and policies regarding deployment to production services and to engage in later patch verification. Moreover, these activities are time-critical given that the longer patches take to be implemented, the more time attackers have to exploit the related vulnerabilities. 

There have been, and continue to be, various examples of attackers using publicized but unpatched vulnerabilities to launch their attacks; even the scanning for these vulnerabilities by attackers continues for periods after patches have been released~\cite{Barracuda2021vul}. This problem has been exacerbated by the advent of generative artificial intelligence (GenAI) systems and Large language models (LLMs) (e.g., ChatGPT, Claude, CoPilot, Gemini, etc.) as they are able to analyze systems for vulnerabilities and even exploit zero-days~\cite{cloudflare2024aivul,fang2024teams,zhou2024large}. Another related and popular example of such scanning is the REvil cybercriminal group who reportedly exploited a vulnerability in Kaseya VSA software at the time that the organization was actively developing a patch~\cite{Ivanti2021vul}. The reality therefore is that patching (quickly) remains an important task for organizations and their security/IT teams as attackers are activity aiming to exploit such weaknesses. This, however, does not always occur, hence our goal of exploring this topic further based on industry and academic literature in the subsequent sections.

\section{The Incentives that Support the Decision to Patch}
\label{sec:incentives}
The question of why security/IT teams make the active decision to patch their organization's systems is a simple, yet complex one. To consider this, we first call attention to the different pertinent stakeholders involved: the organization itself (including IT/security teams), the vendors (inclusive of developers) that it purchases products from, the organization’s business partners and clients, and the legal/regulatory jurisdictions that the enterprise operates in. Each of these can have an impact on decisions regarding patching.

To begin with the organization, patching as an activity is directly in the interest of the IT/security personnel and wider enterprise teams because it helps to secure the company against vulnerability exploits. A primary technique that threat actors use to compromise systems is unpatched vulnerabilities~\cite{NAO2017hos}; therefore, this acts to mitigate related attack vectors. Although cybersecurity may arguably still be intangible and a challenge for many organizations, the financial and reputational impacts/harms that would result from a successful breach~\cite{agrafiotis2018taxonomy,bada2020social,mott2024there}, are salient. This may mean, for instance, that in the interest of self-preservation, concerns about finances or reputational damage can serve as key motivators for companies to addressing security concerns, such as patches. These factors can be central to any proposals that security teams may need to make to their business colleagues if there are internal deliberations about whether to apply a patch. In addition to the security it provides, related benefits to patching have also been cited in minimizing the organization’s downtime in the medium-to-long term, under the assumption that falling victim to a cyber-attack will result in a longer period of unavailable services~\cite{dissanayake2022software}. This, of course, is a delicate argument to make considering that patching itself also often requires downtime; however, the length of downtime and its predictability are likely to be better compared to a security breach. This is an argument that security professionals can use in engaging with their enterprise counterparts as part of a business case for patching for instance.

For IT/security staff, another enabler to the patching of systems is the availability of more efficient and effective processes and tools to allow it. Tools like automated patching~\cite{dissanayake2022empirical}, albeit not perfect, can significantly reduce many of the challenges to patching for security teams; in one study, 76\% of organizations used some form of automated patching and 46\% used live patching~\cite{Tux2021vul}. The wider roll out of these processes and tools may therefore assist IT/security personnel even further in addressing the patch management tasks at hand. Another benefit is that as a precursor to determining which systems are vulnerable, the tools need to find and scan all endpoints on a network. Prior to any patching therefore, this action can already help to detect endpoints and systems that administrators may have previously been unaware of. More generally, our also highlighted that automated patching, for instance, has been promoted due to its ability to allow organizations to better comply with data protection regulations~\cite{Neagu2024vul}. 

Legal and regulatory requirements, business partners, and clients are another significant incentive for patching. Over the last decade, governments have become more stringent regarding the security requirements placed on organizations as evidenced by the EU's General Data Protection Regulation (GDPR), NIS Directives, new US Securities and Exchange Commission (SEC) cybersecurity rules, NIST Cybersecurity Framework, and NCSC’s Cyber Essentials. Many of these regulations and standards have provisions for maintaining the security of systems, which directly point to the need for timely patching. Undoubtedly, for many security teams especially those that work in heavily regulated sectors such as finance and healthcare, the directive to follow these regulations and standards will suffice. In other cases, however, businesses may ignore these either intentionally due to lack of resources, or unintentionally due to lack of awareness, until they have no other choice. For instance, it is noteworthy that despite its wide publicity and it being used as a pre-requisite to apply for some government contracts, the awareness of (and adherence to) the five-control Cyber Essentials scheme is still below 30\% in 2024 in the UK~\cite{DSIT2024cbs}. 

There has also been an increase in international government-initiated efforts to notify organizations about key vulnerabilities and nudge them towards installing patches. One example is the Joint Cybersecurity Advisory in 2023 by the UK’s NCSC, US’ CISA and FBI, and Australia’s ACSC on the top routinely exploited vulnerabilities~\cite{NCSC2023topv}. This advisory complements existing national advisories and alerts issued by these national cybersecurity bodies separately. An noteworthy development pertaining to how governments engage regarding patches was the emergency directive by CISA on the 17 December 2021 ordering federal agencies to investigate and patch their systems against the Log4j vulnerability by the 23 December 2021~\cite{CISA2021emer}. This directive speaks directly to those impacted and requires organizations and their IT/security teams to patch, and report on the update status to CISA (by the 28 December 2021). 

The US’ FTC also subsequently released a warning for organizations to address the vulnerability or face the consequences: ``\textit{The FTC intends to use its full legal authority to pursue companies that fail to take reasonable steps to protect consumer data from exposure as a result of Log4j, or similar known vulnerabilities in the future}''~\cite{FTC2022log}. It also then cited Equifax’s failure to patch, the later impact on 147 million consumers, and the \$700 million settlement, as an example of what can happen to organizations who fail to comply. Such prompts center on cybersecurity teams and business leaders, as they can provide the `push' incentive for leaders to prioritize patching even though it may have a short-term impact on productivity or services. Since that CISA directive, there have been several more with the most recent in 2024 pertaining to mitigating vulnerabilities in software from Ivanti~\cite{CISAIvanti2024emer}. In the UK, similar directives are not as present. However, there are various examples of public reprimands by the Information Commissioner's Office (ICO). One poignant example is the reprimand of the London Borough of Hackney after it's data breach for actions such as the fact that it, ``\textit{failed to ensure that a security patch management system was actively applied to all devices}''\cite{ico2024hackney}. These very public notices seek to nudge organizations---here, local government councils---towards better cybersecurity practices (or face the risk of similar reprimands and potential fines). 

The aforementioned actions by CISA and the FTC seek to mandate certain behavior and will have had an impact on organizations (though there is, at this point, little data available to investigate this). If we consider this discussion more broadly however, there is an argument that these mandates were only released due to the significance of the vulnerability, coupled with the fact that many enterprises may not be aware that they use Log4j and thus may be vulnerable. This is, therefore, unlikely to be standard practice which means that many lower-impact vulnerabilities (be it in number of organizations impacted or exploits possible from the vulnerability) would fail to be directly mandated and instead limited to alert/advisory status if covered at all. Potentially in response to concerns such as this, the ACSC has taken the pre-emptive approach to recommend timeframes for various types of patches. For example, to mitigate advanced cyber threats to internet-facing services, it recommends that patches be installed within two weeks, or within 48 hours if an exploit exists~\cite{acsc2023patching}. This guide directs businesses on best practice (as seen by ACSC) for patching in a simple, straightforward document. A clear report articulated in this way is likely to be well-received given the wider complexity around patching, however, it remains to be seen whether the recommended timescales are always appropriate.

Business partners and clients can influence a company’s decisions to patch, primarily through supply chain agreements (and service-level agreements (SLAs)) and contracts. The rise in supply chain attacks over the last decade has resulted in an increased level of attention to the security of all supply chain parties. Current guidance from bodies such as SANS and insight from academic research suggest that organizations in supply chains require others to implement defined security controls, adhere to security monitoring requirements, and notify of breaches or incidents amongst other actions~\cite{melnyk2022new,SANS2015sc}. It is conceivable therefore that as a part of these contracts and agreements the timely remediation of vulnerabilities via patching is contractually required. On this topic, it is also prudent to mention the influence of other third parties such as business and cyber insurance providers and even financial institutions such as banks/lenders. The influence that cyber insurance companies, in particular, have on the cybersecurity practices of organizations (inclusive of Small-to-Medium-sized Enterprises (SMEs)) has been demonstrated in prior work~\cite{adriko2024does,adriko2024cybersecurity}, and therefore this may be another area where positive pressure may be applied.

Vendors also have a key part to play in encouraging good patch management in organizations. The primary need from their perspective is to provide timely and robust patches, with suitable documentation for organizations to follow. CISA and NIST further suggest that tools should also be provided to support at each stage of the deployment and testing cycle~\cite{CISA2021supp}. This ensures that organizations possess all the guidance needed to remediate vulnerabilities as quickly and effectively as possible. A positive patching experience for an IT/security team can have a significant impact on future decisions to patch; this is also why patches that are slow, difficult to implement, or cause unforeseen problems are so reputationally damaging to the notion of patching. 

Considering the central role of vendors, we also explored research pertaining to how this stakeholder group engaged with the vulnerability disclosure and patching process. In one early study, it was found that vendors are significantly more likely to release patches faster in instances where vulnerabilities are disclosed to them and the public at the same time~\cite{arora2010empirical}. The public disclosure of vulnerabilities is a highly contentious topic, with arguments for it (e.g., it nudges vendors to release of patches quicker) and against it (e.g., it provides attackers with an unnecessary advantage). The study also found that open-source vendors patch faster than closed source organizations (a finding also supported by other work ), and that vendors respond to (patch) vulnerabilities quicker if they are disclosed by CERT (or similar authoritative institutions). The latter of these points is noteworthy and highlights the key role that such bodies play in nudging vendors to release faster (and ideally higher quality) patches which then influences how organizations engage with (e.g., the trust they place in, the speed of implementation of) their own internal patching processes. 

Other research makes a similar point and notes that when there are legislative pressures, vendors patch vulnerabilities faster~\cite{temizkan2012patch}. This is significant for legislators and those that influence policy---a key caveat nonetheless is that a quicker release of a low-quality patch may also have a substantial negative impact. In considering these findings, readers should note that the two studies above are both quite dated and may not represent the current state of play; unfortunately, these articles are the most recent academic work that we could find investigating this specific topic. This clearly represents a gap in current research that should be addressed.

Finally, some literature has considered how to motivate individuals (i.e., the ultimate end users) to install updates and patches. While this is not based directly at the organizational level, it is within the remit of human aspects and given that non-IT staff in organizations have an influence on whether a patch is installed or not on their local devices, they may refuse to install updates. Recommendations target various areas including suggesting that updates/patches be installed at convenient times (e.g., over lunch), an emphasis on education, transparency and communication, and highlighting the benefits/value~\cite{ftc2019encourage,ftc2021action}. They also consider more punitive approaches such as revoking access of those who do not update, making activities/support difficult (which may also include forcing updates before the system or application can be used), rating departments based on update status, and refusing to patch old systems (thus making older systems less attractive for use, and encourage staff to install newer systems). These approaches all have their pros and cons and will need to acknowledge the nuances of end users, their needs and context, and the wider organizational (security) culture~\cite{uchendu2021developing}. The reality is that although corporate policy typically mandates certain processes, there are always edge cases that need to be accounted for, be it the CEO of an organization or a user who has found a way to circumvent patching in the interest of being more efficient at their work.

\section{What Motivates the Decision Not to Patch?}
\label{sec:disincentives}

Patching by its nature is a challenging task which often involves a series of interdependent activities, complex networks, a range of applications, business processes and people. Yet, patching is critical to cybersecurity and failure to patch can result in breaches. The data breach of Equifax in 2017 which impacted 147 million individuals provides an poignant case study, which also highlights the human aspects involved~\cite{USS2017eq}. In their case, organizational teams failed to define strategies for patch management, failed to act on vulnerabilities and install available patches, failed to define their assets (a prerequisite to understanding vulnerabilities that may be present), and failed to specify and follow a reliable process to gather vulnerability and patch information from product vendors---in this case, it relied on individual software developers. Such failings are, however, not new and arguably are as a result of the challenging nature of managing and deploying patches. Other more recent examples of organizations that have not patched and suffered breaches include the LastPass, Electoral Commission and the London Borough of Hackney~\cite{ico2024electoral,ico2024hackney,thn2023lp}.

There are at least five primary reasons why vulnerabilities are often not patched~\cite{NCSC2019pat}. The first reason considers the resources needed to patch systems, e.g., the money, time and person-power required. For some IT/security teams, resources may not exist to support patch management, or their environments may be so complex that patch management is a minefield. If we consider micro businesses or Small-to-Medium-sized Enterprises (SMEs) as an example, their priorities are often elsewhere, and they are known to struggle with security generally even lacking dedicated IT/security teams~\cite{bada2019developing,khan2024assessing}. Even if such organizations want to patch, they may not have the internal skills or expertise to test, implement and deploy it, or they may not be able to patch in a timely manner. A best-case scenario in this context is that internal teams work with managed (security) service provides or adopt software-as-a-service platforms; and that these providers then handle patching for the SME. This, however, is not always the case.

For larger organizations and IT/security teams who have the person-power, finances and time, the challenge instead is often on the alignment of these resources to identify, implement and follow-up patch management tasks in their extremely complex environments. Equifax is a relevant example of this where there were numerous mistakes and failures in the process, and there was little accountability demonstrated. Further evidence can be seen in the statement made by their Chief Information Officer who led the IT department during 2017, referring to patching as a “lower level responsibility that was six levels down” from him~\cite{USS2017eq}. The size and complexity of the organization clearly impacted how significant issues such as patching were perceived. This resonates with existing studies which discovered that over 70\% of security professionals found patching to be too complex and time consuming~\cite{Ivanti2021pat}. 

Another major challenge facing both small and large organizations and their IT/security teams is the volume of patches that need to constantly be applied and the question of what to patch first in such situations. These issues have been identified within the top four reasons why organizations delay patches~\cite{SerPon2019cost}. While automated patching tools can help the process, they are not a panacea. A telling statistic from another industry study was that 72\% of respondents did not apply patches because of the difficulty in prioritizing what needs to be patched~\cite{SerPon2019cost}. While scoring systems such as CVSS exist and are used to prioritize vulnerabilities, the reality is that internal teams will will need to pair these with the impact on an organization’s individual systems/assets to judge the risk exposure. This additional step can also prove difficult depending on the internal team's understanding of their own environment. A clear understanding of IT, security and business dependencies as well as close relationships between these teams is therefore also paramount. 

The second reason relates to the challenge of knowing an organization's full portfolio of assets and their state. This links to the complexity of organizations and the reality that maintaining an up-to-date understanding of all assets on a corporate network is a significant undertaking for IT/security teams. Organizations can possess thousands of networked devices, systems and applications across IT, OT and IoT environments. Siloed systems and dynamic networks have also been recognized directly as key contributors to this complexity~\cite{SCMag2021as}. The widespread move to remote working due to COVID-19 travel and social restrictions has made these networks even more complicated, which has further exacerbated the scale of the patch management problem~\cite{Ivanti2021pat}. With these points in mind, it is unsurprising that cybersecurity schemes such as CIS's Critical Security Controls (CSC) spotlight `Inventory and Control of Enterprise Assets' and `Inventory and Control of Software Assets' as the first two controls that are central to organizational security. 

The third and fourth reasons acknowledge the risk that accompanies patching and the significant harm that may be caused if patches negatively impact existing systems and inadvertently cause business disruption. The reason why systems may be adversely impacted can vary and pertain to issues at the vendor (e.g., low quality or untested patch) or the IT/security product owner (e.g., failure to implement patch as directed, or absence of other key prerequisite patches or bug fixes). A notable example of a case where a security patch resulted in significant disruption is the CrowdStrike patch incident in 2024. Here, a defective patch released by CrowdStrike inadvertently crashed computers across the world and led to unavailable systems at hospitals, airlines, banks and other businesses; estimates suggest that it costed \$5.4bn in the US alone~\cite{theguardian2024crowd}. Separately, although we were unable to find the specific reason why NHS systems were not patched and thus were vulnerable to WannaCry~\cite{NAO2017hos}, it is not inconceivable that concerns regarding system availability---given the criticality of their systems---factored into such a decision. Other studies support this general idea, with some surprisingly reporting that over half of professionals surveyed believe that patches (when not properly executed) cause greater risk of instability than a data breach~\cite{Pon2018myt}. 

Inadvertently disrupting systems is a vital concern, but another one is the reality that some patches may fail to address the series of related vulnerabilities completely, thus requiring further patches. For security teams therefore, this would mean preparing, testing, installing, and deploying patches for related vulnerabilities multiple times. Such eventualities pose clear disincentives for these teams to want to patch, or to patch in a timely manner. They also impact trust in the vendor and the patching process.

The final reason is that some systems may not be patchable. This may be due to the vendor no longer publishing updates, the specialized nature of the product (e.g., medical devices), or responsibility for the product, or its patching, existing elsewhere. Medical devices and other similarly critical products pose a particularly difficult challenge regarding patching. These devices are, as to be expected, thoroughly checked and vetted prior to their sale. However, patching remote products deployed in various types and conditions of environment can be uniquely perplexing. In the UK, NHS Digital’s guidance, for instance, notes that patching medical devices can take three months from the time that a security patch is available~\cite{NHSDig2022guid}; even this ideal situation presents a clear window for attackers in a critical national infrastructure service. 

Beyond the primary factors reviewed above, research provides some further insights which complement the points above but also allude to others including challenges with manual patch management tasks, human error, and the perception that vulnerabilities would not be exploited~\cite{SerPon2019cost,Tux2021vul}. The perception that attackers will not exploit vulnerabilities arguably relates to the intangibility of cyber risk particularly for those not in security teams. However, there is also some truth to this point as discovered by other work which suggests that the majority of registered vulnerabilities are not exploited~\cite{Nather2018pat}. The position therefore is that inaction may be a plausible strategy---less risky than deploying (potentially problematic) patches---assuming that attackers are unlikely to find the vulnerability in that specific organization. 

A related point to those covered above is that of patch management and monitoring tools; these can support IT/security teams significantly in addressing vulnerabilities quickly. This is especially relevant now that artificial intelligence (e.g., GenAI) is being explored more to support the patching process~\cite{akuthota2023vulnerability,liu2025enhancing,liu2024exploring,martinez2025generative}. Unfortunately, they themselves may also pose a source of problems. For instance, it has been noted that patch management tools can give conflicting reports regarding the status of a patch, i.e., one tool states that a system is patched and another reports that it is not~\cite{Alexiou2019pat}. Additionally, these tools themselves may be complex to run and even platform specific. This confusion and difficulty is likely to frustrate IT/security teams and discourage them from adequately engaging with the patch management process. Compatibility of tools can also lead to other issues. Automated agents commonly used to provide endpoint protection (e.g., on systems that cannot be patched for some reason) are not always allowed to execute on vendor systems; moreover, running these agents may void support or warranty agreements~\cite{Alexiou2019pat}. Such interactions been vendors (IT and security) only succeeds in harming the organization, and potentially forcing staff not to patch systems for fear of further negative business or supplier impact. 

The final salient point relates again to vendors and the fact that supply chain attacks on vendors can further act against the adoption of good patch management practices. The attack on SolarWinds, for example, was significant, but especially so because of how a vendor (i.e., SolarWinds) that supplied several critical organizations (e.g., various businesses worldwide including telecommunications organizations, US Fortune 500 companies, the US Military, and the Pentagon) was breached, and then used to compromise clients via system updates~\cite{CSO2020sw}. This type of situation now presents a significant dilemma, where organizations and IT/security teams should install patches because it is security best practice but may simultaneously be worried about whether the patches are legitimate (or, at an extreme, contain malware). This questions the trust placed in patches and will undoubtedly add to the various other concerns that may lead to deciding against patching in a timely manner.

\section{Discussion and Future Considerations}
\label{sec:reflection}

To maintain the security of organizations and systems, it is a critical activity for IT/security teams to install security updates. The incentives for patching are varied and include fulfilling legal and regulatory requirements or security-related requests from business partners, protecting the organization from security incidents, the availability of efficient and effective tools to support the patch management process for security teams, and a positive patching experience for IT/security personnel, especially as it relates to vendor engagement. Similarly, there are several reasons why organizations do not, or are disincentivized, to patch their systems. Our review detailed issues related to the nontrivial resources (e.g., person-power) needed to implement patches, the complexity of modern-day technology environments, the volume of patches to apply (and all of varied priority levels), the risk of inadvertent business interruption, and key human aspects factors such as reservations about tools meant to support patch management, the perception by companies (and evidence that suggests) that vulnerabilities are not often exploited, and potential questions about trust in vendors. 

Comparing the incentives and disincentives there are compelling reasons on both sides. Legal and regulatory requirements are clearly one of the strongest ways to nudge organizations to install patches, and in a timely manner. If the regulation exists, then IT/security teams (driven by corporate guidance) will have to, in theory, find the resources to overcome issues of resourcing and complexity to implement the requisite patches. This may, however, be considered a more heavy-handed approach  and governments may not want to take this route in a general case, but rather use actions such as mandates for extremely critical vulnerabilities, such as with the situation with Log4J and CISA. There is also the innate challenge with mandating any security controls or actions for all of a nation’s organizations and how this would be actively promoted, checked and enforced. There have been interesting discussions for several years on the role of cyber insurance in security  and whether this may be a route to incentivize better security practices; this, however, has yet to materialize~\cite{adriko2024does,maccoll2021cyber}. Business partners are another strong avenue to incentivise patching if such partners are powerful enough to make demands of others in the supply chain, and if they fully appreciate the threat posed by unpatched partner systems themselves. The argument therefore is that if key supply chain organizations were successfully engaged, then they could place certain requirements related to patching into their service and security agreements with others in their chains.

While there is clearly a strong argument that organizations interested in self-preservation will install patches (and that this itself, should suffice for motivation), this may be directly contested by concerns about the human resources (time, person-power, technical expertise) needed to implement patches, the complexity of enterprise environments (where patches are to be applied), the potential that patches may disrupt operations (even if it is linked solely to poor execution or human error), and the perception and reality that only a small number of vulnerabilities are actively exploited in the wild. Even in cases where installing patches from a vendor may be possible, mitigation approaches (e.g., virtual patching) may be pursued by (already overworked) security teams instead because they are less invasive. Addressing these concerns is difficult and requires efforts from organizations and vendors. One way forward may be improved tooling that is capable of more effective and efficient automation of patch management tasks. This could reduce the workload on IT/security teams, both in understanding the interdependence of affected systems and implementing required security updates. Automated tools---even those built on generative AI systems and LLMs~\cite{akuthota2023vulnerability,liu2024exploring}---have already been highlighted as an enabler to patch management, but challenges also exist (e.g., conflicting reports, vendor stipulations) hence the need for improvements. These challenges make the role of security teams extremely difficult, and at a time where the industry is facing serious issues with respect to burnout~\cite{reeves2023your}. 

Vendors also play a central role in either enabling or complicating the patch management activities in organizations. As identified from our review, a positive patching experience can have a notable impact on an IT/security (or business) team's desire (or support) to patch. This would therefore suggest that if vendors were to invest more resources into supporting security updates, therefore producing higher quality patches---as well as patch guidance and deployment frameworks---in a quicker manner, their uptake may be increased. A notable development here is the OASIS Common Security Advisory Framework (CSAF) which although it relates to vulnerability notification is important as it demonstrates a move by industry towards better communication system to engage with product owners. Any efforts by vendors to support patch management can help to address these concerns and others, including those about the lack of trust product owners have in vendors or concerns about the unintentional business risks posed by patches. 

Although the challenges to producing and releasing timely, high-quality patches is clearly nontrivial, it is an important requirement to build trust in the patching process. We also stress that as supply chain attacks become more rampant, vendors need to increase their protection mechanisms and avoid scenarios where breaches in vendor systems can subsequently result in malware deployment to customers via patches and updates. At the very least, such attacks (or the news of such attacks) will cause internal teams to take longer to install patches (while they conduct additional testing and vetting), or in the worst case, result in some teams recommending against patches completely (or opting for complex mitigations that do not resolve the core problem). 

To complement the discussions on incentives and disincentives, there has been some relevant work directed specifically at solutions to these intertwined challenges. For instance, one suggestion has been that when owners or IT staff within small-to-medium enterprises (SMEs) look to purchase a software product, they should first check if it is available as a service~\cite{Culafi2021pat}. If available, the service offering should then be preferred, presumably because this places the responsibility on patching with the provider which may increase the chances of a swifter implementation of patches. A similar argument can be made for enterprises making the use of more cloud computing services given that this may have the same impact. NIST’s SP 800-40 adds further weight to these points as it suggests actions such as using managed services instead of software when feasible, and working with providers who are less likely to create poor, vulnerability-prone software~\cite{NIST2022guidepat}. These could be reasonable options for security teams assuming that reliance on a third-party does not led to further, or other problems.

\section{Conclusion}
\label{sec:conclusion}

Motivating organizations and their IT/security teams to install security patches on their systems has been a problem for decades. As discussed in this article, there are various reasons for this reticence but also several factors that highlight the value of patching and that support the process. If the uptake of patching is to be increased,  key incentive areas should be further emphasized. In particular, we would suggest enhancing tooling and automation leveraging generative AI platforms to support the work of cybersecurity staff, better support from and engagement with vendors, security nudges to organizations leveraged via trusted supply chain partners, increased pressure on vendors to deliver products which are less vulnerable while also producing patches of higher quality, and finally if the situation calls for it, appropriate regulation. There may also be the opportunity for new systems development paradigms to gain in popularity such as evergreen IT, with its incremental, iterative updates; thus reducing the likelihood of particularly vulnerable, dated IT systems. There would also be less work for security teams which helps with workload and other human aspects issues. In either case however, while patching may be considered by some as a failing paradigm~\cite{Schneier2018pat}, the reality is that currently it is arguably impossible to create a perfectly secure system and therefore patching in some form is likely here to stay.

%
%
%
%
\bibliographystyle{splncs04}
\bibliography{bib}

\begin{thebibliography}{10}
\providecommand{\url}[1]{\texttt{#1}}
\providecommand{\urlprefix}{URL }
\providecommand{\doi}[1]{https://doi.org/#1}

\bibitem{Abbasi2023vul}
Abbasi, S.: {2023 Threat Landscape Year in Review: If Everything Is Critical, Nothing Is} (2023), {https://blog.qualys.com/vulnerabilities-threat-research/2023/12/19/2023-threat-landscape-year-in-review-part-one}

\bibitem{acsc2023patching}
{ACSC}: {Patching Applications and Operating Systems} (2023), {https://www.cyber.gov.au/resources-business-and-government/maintaining-devices-and-systems/system-hardening-and-administration/system-administration/patching-applications-and-operating-systems}

\bibitem{adriko2024cybersecurity}
Adriko, R., Nurse, J.R.C.: {Cybersecurity, cyber insurance and small-to-medium-sized enterprises: a systematic Review}. Information \& Computer Security  \textbf{32}(5),  691--710 (2024)

\bibitem{adriko2024does}
Adriko, R., Nurse, J.R.C.: {Does Cyber Insurance promote Cyber Security Best Practice? An Analysis based on Insurance Application Forms}. Digital Threats: Research and Practice  \textbf{5}(3),  1--39 (2024)

\bibitem{agrafiotis2018taxonomy}
Agrafiotis, I., Nurse, J.R.C., Goldsmith, M., Creese, S., Upton, D.: A taxonomy of cyber-harms: Defining the impacts of cyber-attacks and understanding how they propagate. Journal of Cybersecurity  \textbf{4}(1) (2018)

\bibitem{akuthota2023vulnerability}
Akuthota, V., Kasula, R., Sumona, S.T., Mohiuddin, M., Reza, M.T., Rahman, M.M.: Vulnerability detection and monitoring using llm. In: IEEE 9th International Women in Engineering (WIE) Conference on Electrical and Computer Engineering (WIECON-ECE). pp. 309--314. IEEE (2023)

\bibitem{Alexiou2019pat}
Alexiou, S.: Practical patch management and mitigation. ISACA Journal  \textbf{3} (2019)

\bibitem{arora2010empirical}
Arora, A., Krishnan, R., Telang, R., Yang, Y.: An empirical analysis of software vendors' patch release behavior: impact of vulnerability disclosure. Information Systems Research  \textbf{21}(1),  115--132 (2010)

\bibitem{bada2019developing}
Bada, M., Nurse, J.R.C.: {Developing cybersecurity education and awareness programmes for small-and medium-sized enterprises (SMEs)}. Information \& Computer Security  \textbf{27}(3),  393--410 (2019)

\bibitem{bada2020social}
Bada, M., Nurse, J.R.C.: {The Social and Psychological Impact of Cyberattacks}. In: Emerging cyber threats and cognitive vulnerabilities, pp. 73--92. Elsevier (2020)

\bibitem{Barracuda2021vul}
{Barracuda}: {Threat Spotlight: Unpatched software vulnerabilities} (2021), {https://blog.barracuda.com/2021/07/21/threat-spotlight-unpatched-software-vulnerabilities/}

\bibitem{bbc2024crowd}
{BBC News}: {CrowdStrike IT outage affected 8.5 million Windows devices, Microsoft says} (2024), {https://www.bbc.co.uk/news/articles/cpe3zgznwjno}

\bibitem{CISA2021supp}
{CISA}: {Defending Against Software Supply Chain Attacks} (2021), {https://www.cisa.gov/sites/default/files/publications/defending\_against\_software \_supply\_chain\_attacks\_508\_1.pdf}

\bibitem{CISA2021emer}
{CISA}: {Emergency Directive 22-02, “Mitigate Apache Log4j Vulnerability”} (2021), {https://www.cisa.gov/emergency-directive-22-02}

\bibitem{CISAIvanti2024emer}
{CISA}: {Supplemental Direction V2: ED 24-01: Mitigate Ivanti Connect Secure and Ivanti Policy Secure Vulnerabilities} (2024), {https://www.cisa.gov/news-events/directives/supplemental-direction-v2-ed-24-01-mitigate-ivanti-connect-secure-and-ivanti-policy-secure}

\bibitem{cisa2025cat}
{CISA}: {Known Exploited Vulnerabilities Catalog} (2025), {https://www.cisa.gov/known-exploited-vulnerabilities-catalog}

\bibitem{cloudflare2024aivul}
{Cloudflare}: {Can AI find vulnerabilities?} (2024), {https://www.cloudflare.com/en-gb/the-net/ai-vulnerabilities/}

\bibitem{CSO2020sw}
{CSO}: {SolarWinds attack explained: And why it was so hard to detect} (2020), {https://www.csoonline.com/article/3601508/solarwinds-supply-chain-attack-explained-why-organizations-were-not-prepared.html}

\bibitem{Culafi2021pat}
Culafi, A.: {Why patching vulnerabilities is still a problem, and how to fix it} (2021), {https://www.techtarget.com/searchsecurity/news/252503950/Why-patching-vulnerabilities-is-still-a-problem-and-how-to-fix-it}

\bibitem{dissanayake2022software}
Dissanayake, N., Jayatilaka, A., Zahedi, M., Babar, M.A.: Software security patch management-a systematic literature review of challenges, approaches, tools and practices. Information and Software Technology  \textbf{144},  106771 (2022)

\bibitem{dissanayake2022empirical}
Dissanayake, N., Jayatilaka, A., Zahedi, M., Babar, M.A.: An empirical study of automation in software security patch management. In: The 37th IEEE/ACM International Conference on Automated Software Engineering. pp. 1--13 (2022)

\bibitem{DSIT2024cbs}
{DSIT}: {Cyber Security Breaches Survey 2024} (2024), {https://www.gov.uk/government/statistics/cyber-security-breaches-survey-2024/cyber-security-breaches-survey-2024}

\bibitem{ENISA2018sw}
{ENISA}: {Is Software More Vulnerable Today?} (2018), {https://www.enisa.europa.eu/publications/info-notes/is-software-more-vulnerable-today}

\bibitem{fang2024teams}
Fang, R., Bindu, R., Gupta, A., Zhan, Q., Kang, D.: {Teams of LLM Agents can Exploit Zero-Day Vulnerabilities}. arXiv preprint arXiv:2406.01637  (2024)

\bibitem{ftc2019encourage}
{Forbes Technology Council}: {Encourage Your Team To Install Software Updates With These 10 Tactics} (2019), {https://www.forbes.com/sites/forbestechcouncil/2019/01/30/encourage-your-team-to-install-software-updates-with-these-10-tactics/}

\bibitem{ftc2021action}
{Forbes Technology Council}: {Actionable Strategies To Get Your Software Users To Install Updates} (2021), {https://www.forbes.com/sites/forbestechcouncil/2021/02/05/14-actionable-strategies-to-get-your-software-users-to-install-updates/}

\bibitem{frei2010modeling}
Frei, S., Schatzmann, D., Plattner, B., Trammell, B.: Modeling the security ecosystem-the dynamics of (in) security. In: Economics of Information Security and Privacy, pp. 79--106. Springer (2010)

\bibitem{FTC2022log}
{FTC}: {FTC warns companies to remediate Log4j security vulnerability} (2022), {https://www.ftc.gov/news-events/blogs/techftc/2022/01/ftc-warns-companies-remediate-log4j-security-vulnerability}

\bibitem{SCMag2021as}
Guy, J.: {Asset inventory has become a serious security problem} (2021), {https://www.scmagazine.com/perspective/asset-inventory-has-become-a-serious-security-problem}

\bibitem{ico2024electoral}
{ICO}: {ICO reprimands the Electoral Commission after cyber attack compromises servers} (2024), {https://ico.org.uk/about-the-ico/media-centre/news-and-blogs/2024/07/ico-reprimands-the-electoral-commission-after-cyber-attack-compromises-servers/}

\bibitem{ico2024hackney}
{ICO}: {London Borough of Hackney reprimanded following cyber-attack} (2024), {https://ico.org.uk/about-the-ico/media-centre/news-and-blogs/2024/07/london-borough-of-hackney-reprimanded-following-cyber-attack/}

\bibitem{Ivanti2021vul}
{Ivanti}: {Ransomware Index Spotlight Report Reveals Steady Increase in Sophistication and Volume of New Ransomware Vulnerabilities and Families} (2021), {https://www.ivanti.co.uk/company/press-releases/2021/ransomware-index-spotlight-report-reveals-steady-increase-in-sophistication-and-volume-of-new-ransomware-vulnerabilities-and-families-in-q3-2021}

\bibitem{Ivanti2021pat}
{Ivanti}: {Patch Management Challenges} (2024), {https://www.ivanti.com/resources/v/doc/ivi/2634/712cff539c8a}

\bibitem{khan2024assessing}
Khan, N., Furnell, S., Bada, M., Nurse, J.R.C., Rand, M.: Assessing cyber security support for small and medium-sized enterprises. In: International Symposium on Human Aspects of Information Security and Assurance. pp. 148--162. Springer (2024)

\bibitem{thn2023lp}
Lakshmanan, R.: {LastPass Hack: Engineer's Failure to Update Plex Software Led to Massive Data Breach} (2023), {https://thehackernews.com/2023/03/lastpass-hack-engineers-failure-to.html}

\bibitem{liu2025enhancing}
Liu, J., Lin, G., Mei, H., Yang, F., Tai, Y.: Enhancing vulnerability detection efficiency: An exploration of light-weight llms with hybrid code features. Journal of Information Security and Applications  \textbf{88},  103925 (2025)

\bibitem{liu2024exploring}
Liu, P., Liu, J., Fu, L., Lu, K., Xia, Y., Zhang, X., Chen, W., Weng, H., Ji, S., Wang, W.: {Exploring ChatGPT's Capabilities on Vulnerability Management}. In: 33rd USENIX Security Symposium. pp. 811--828 (2024)

\bibitem{maccoll2024ransomware}
MacColl, J., H{\"u}sch, P., Mott, G., Sullivan, J., Nurse, J.R., Turner, S., Pattnaik, N.: Ransomware: Victim insights on harms to individuals, organisations and society  (2024)

\bibitem{maccoll2021cyber}
MacColl, J., Nurse, J.R.C., Sullivan, J.: Cyber insurance and the cyber security challenge. RUSI Occasional Paper  (2021)

\bibitem{martinez2025generative}
Mart{\'\i}nez, A.L., Cano, A., Ruiz-Mart{\'\i}nez, A.: Generative artificial intelligence-supported pentesting: A comparison between claude opus, gpt-4, and copilot. arXiv preprint arXiv:2501.06963  (2025)

\bibitem{melnyk2022new}
Melnyk, S.A., Schoenherr, T., Speier-Pero, C., Peters, C., Chang, J.F., Friday, D.: New challenges in supply chain management: cybersecurity across the supply chain. International Journal of Production Research  \textbf{60}(1),  162--183 (2022)

\bibitem{mott2024there}
Mott, G., Turner, S., Nurse, J.R.C., Pattnaik, N., MacColl, J., Huesch, P., Sullivan, J.: {‘There was a bit of PTSD every time I walked through the office door’: Ransomware harms and the factors that influence the victim organization’s experience}. Journal of Cybersecurity  \textbf{10}(1) (2024)

\bibitem{mott2023between}
Mott, G., Turner, S., Nurse, J.R., MacColl, J., Sullivan, J., Cartwright, A., Cartwright, E.: Between a rock and a hard (ening) place: Cyber insurance in the ransomware era. Computers \& Security  \textbf{128},  103162 (2023)

\bibitem{NAO2017hos}
{NAO}: {Investigation: WannaCry cyber attack and the NHS} (2017), {https://www.nao.org.uk/reports/investigation-wannacry-cyber-attack-and-the-nhs/}

\bibitem{Nather2018pat}
Nather, W.: {Patching All The Things May Not Be The Best Strategy} (2018), {https://duo.com/decipher/patching-all-the-things-may-not-be-the-best-strategy}

\bibitem{NCSC2019pat}
{NCSC}: {The problems with patching} (2019), {https://www.ncsc.gov.uk/blog-post/the-problems-with-patching}

\bibitem{NCSC2023topv}
{NCSC}: {NCSC and allies reveal most common cyber vulnerabilities exploited in 2022} (2023), {https://www.ncsc.gov.uk/news/ncsc-allies-reveal-2022-common-exploited-vulnerabilities}

\bibitem{NCSC2024vul}
{NCSC}: {NCSC: Vulnerabilities} (2024), {https://www.ncsc.gov.uk/section/advice-guidance/all-topics?topics=Vulnerabilities}

\bibitem{Neagu2024vul}
Neagu, C.: {Automated Patch Management Explained} (2024), {https://heimdalsecurity.com/blog/automated-patch-management-process/}

\bibitem{NHSDig2022guid}
{NHS Digital}: {Guidance on protecting medical devices} (2022), {https://digital.nhs.uk/cyber-and-data-security/guidance-and-assurance/guidance-on-protecting-connected-medical-devices}

\bibitem{NIST2022guidepat}
{NIST}: {Guide to Enterprise Patch Management Planning} (2022), {https://nvlpubs.nist.gov/nistpubs/SpecialPublications/NIST.SP.800-40r4.pdf}

\bibitem{perez2024patchy}
P{\'e}rez, S.R., van Eeten, M., Ga{\~n}{\'a}n, C.H.: Patchy performance? uncovering the vulnerability management practices of iot-centric vendors. In: 2024 IEEE Symposium on Security and Privacy (SP). pp. 154--154. IEEE Computer Society (2024)

\bibitem{Pon2018myt}
{Ponemon Institute}: {Separating the Truths from the Myths in Cybersecurity} (2018), {https://www.bmc.com/content/dam/bmc/collateral/third-party/Ponemon\%2BReport.pdf}

\bibitem{reeves2023your}
Reeves, A., Pattinson, M., Butavicius, M.: {Is Your CISO Burnt Out yet? Examining Demographic Differences in Workplace Burnout Amongst Cyber Security Professionals}. In: International Symposium on Human Aspects of Information Security and Assurance. pp. 225--236. Springer (2023)

\bibitem{SANS2015sc}
{SANS}: {Combatting Cyber Risks in the Supply Chain} (2015), {https://www.sans.org/webcasts/combatting-cyber-risks-supply-chain-100657/}

\bibitem{Schneier2018pat}
Schneier, B.: {Patching Is Failing as a Security Paradigm} (2018), {https://www.vice.com/en/article/439wbw/patching-is-failing-as-a-security-paradigm}

\bibitem{SerPon2019cost}
{ServiceNow \& Ponemon}: {Costs and Consequences of Gaps in Vulnerability Response} (2019), {https://www.servicenow.com/lpayr/ponemon-vulnerability-survey.html}

\bibitem{sowka2025towards}
Sowka, K., Palade, V., Jiang, X., Jadidbonab, H.: Towards the generation of hierarchical attack models from cybersecurity vulnerabilities using language models. Applied Soft Computing p. 112745 (2025)

\bibitem{temizkan2012patch}
Temizkan, O., Kumar, R.L., Park, S., Subramaniam, C.: Patch release behaviors of software vendors in response to vulnerabilities: An empirical analysis. Journal of management information systems  \textbf{28}(4),  305--338 (2012)

\bibitem{theguardian2024crowd}
{The Guardian}: {CrowdStrike global outage to cost US Fortune 500 companies \$5.4bn} (2024), {https://www.theguardian.com/technology/article/2024/jul/24/crowdstrike-outage-companies-cost}

\bibitem{Tux2021vul}
{Tux Care}: {State of Enterprise Vulnerability Detection and Patch Management} (2021), {https://tuxcare.com/state-of-enterprise-vulnerability-detection-and-patch-management-report/}

\bibitem{uchendu2021developing}
Uchendu, B., Nurse, J.R.C., Bada, M., Furnell, S.: Developing a cyber security culture: Current practices and future needs. Computers \& Security  \textbf{109},  102387 (2021)

\bibitem{USS2017eq}
{United States Senate}: {How Equifax Neglected Cybersecurity and Suffered a Devastating Data Breach: Staff Report} (2017), {https://www.hsgac.senate.gov/imo/media/doc/FINAL\%20Equifax\%20Report.pdf}

\bibitem{zhou2024large}
Zhou, X., Cao, S., Sun, X., Lo, D.: Large language model for vulnerability detection and repair: Literature review and the road ahead. ACM Transactions on Software Engineering and Methodology  (2024)

\end{thebibliography}
\end{document}